# Thermal Shock Effects Modeling on a Globe Valve Body-Bonnet Bolted Flange Joint

Jean-Philippe MATHIEU, EDF R&D, Moret-sur-Loing
Jean-François RIT, EDF R&D, Moret-sur-Loing
Jérôme FERRARI, EDF R&D, Moret-sur-Loing
David HERSANT, EDF R&D, Moret-sur-Loing

**Abstract :**
This paper attends to show efforts made at EDF R&D to improve comprehension of valve parts loadings during operation. Thermal shock in a globe valve is represented and modeled using EDF R&D Finite Element Analysis code (Code_Aster). Choices of modeling are discussed and balanced on the basis of "what an engineer can obtain without becoming a researcher".

First simulation results are presented. Attention is focused on the evolution of Body-Bonnet Bolted Flange Joint (BBBFJ) tightening forces which are simulated during the thermal shock.

An experimental setup is also presented for the studied valve, which implies thermocouple implementation for comparison of the simulated thermal field and strain measurement on each threaded rod to validate the mechanical modeling.

# Modélisation de l'Effet d'un Choc Thermique sur le Joint Corps-Chapeau d'un Robinet à Soupape

**Abstract :**
Cet article montre les actions entreprises à EDF R&D pour améliorer la compréhension des chargements vus par les composants en cours d'utilisation. Un choc thermique dans un robinet à soupape est représenté et modélisé grâce au code élément-fini d'EDF R&D (Code_Aster). On discutera notamment de quelques uns des choix de modélisation, que l'on mettra en regard de « ce que peut faire un ingénieur sans devenir un chercheur ».

Les premiers résultats de simulation numérique sont présentés. On s'intéresse tout particulièrement à l'évolution des efforts de serrage des goujons au joint corps-chapeau simulés lors d'un choc thermique.

On évoquera par ailleurs le dispositif expérimental associé, qui regroupe l'implémentation de thermocouples et des mesures de déformation sur chacun des goujons de la jonction boulonnée, respectivement pour valider la modélisation thermohydraulique et le modèle mécanique.

# 1 Introduction

## 1.1 Context in French nuclear industry

Nuclear industry is certainly to be considered as a specific industry because of some unusual features. Safety considerations and their priority are common with other industries, but some other specificities arise. Specific nuclear standards and long life cycles are required and are sources of issues to suppliers and users. This is also the case when it comes to valve reliability.

In French nuclear industry, EDF used to work closely with valve suppliers to ensure the reliability. Thus, valves where evaluated in the testing loops available at EDF R&D Moret-sur-Loing (France) facility, and then "qualified" by EDF's specific engineering teams, mostly at the company's own expense. This process of qualification was created in the beginning of the French nuclear Pressurized Water Reactors (PWR) industry and helped reduce reactor unavailability due to valve failures. Since industrials are now considered to have acquired sufficient experience during this period, EDF acquires valves differently.

In the current new qualification process, manufacturers of valves are contractually asked by EDF to prove that their components meets the doctrine and its specifications. EDF still keeps some test facilities, reserved to the R&D actions and exceptional testing, including collaborations with suppliers as well as commercial testing [1].

Numerical simulations have also made great progress during this period, but to the author's knowledge, no manufacturer had produced any "numerical functional test" to demonstrate valve reliability yet. This is certainly due to a lack of industrial standards in numerical modeling, which is definitely a great and "difficult to open" lock. More prosaically, the fact that modeling a valve is still an R&D field [2] also prevents its generalization.

These kinds of modeling are generally used at EDF in order to analyze focused aspects of components failures. The NICODEME and, more recently NICODEME II projects [2] were set up at EDF R&D to evaluate the ability of numerical modeling of valve components to (i) validate one technological function (e.g. sealing of body-bonnet) or (ii) validate a manufacturer argumentation (e.g. qualification of different main diameters of the same valve once one has been experimentally qualified), (iii) validate a valve design. Presently, research effort mainly concerns the two first aspects.

The aim of this paper is to discuss the first of these three aspect, and more especially the threaded rods of the BBBFJ tightening behavior in a globe valve during thermal shock.

## 1.2 Numerical modeling and valve reliability assessment during thermal shock

As far as author knows, few papers deal with a complete realistic numerical simulation of a thermal shock inside a valve of the primary circuitry of a nuclear reactor. This can be explained by the fact that these simulation are difficult to setup, hard to validate experimentally, and eventually require good engineering skills to interpret.

Recent publications addressed issues about designing of valves used in Liquefied Natural Gas (LNG). Sugita investigated different aspects of valve design with respect to sealing performance in cryogenic condition [3], whereas Ravindran et al. exposed some results about thermal transient shock effects inside valves [4].

It is worth noting that the frontier between the two worlds (design on the supplier's point of view, and validation on the customer's one) is not strictly marked : the same problem arises. Once you make numerical simulations of valve components under such extreme loadings (cryogenic fluids, PWR primary circuitry…), it is quite frequent that stresses cross the line of "yield strength". This implies that more and more material and fracture mechanics are needed to give a valid diagnostic (moreover when dealing with some exotic material like packing [5] or hard-facing alloys). Manufacturing related residual stresses are also to be taken into account, or at least cleared of any bad effect.

The first step of NICODEME demonstrated the feasibility of a simulation of a thermal shock inside a globe valve, with an experimental validation of the thermal fields near flow . Because of difficulties previously exposed, it was chosen in this part to study a part of a valve that would allow to validate the mechanical aspects also, but which allows to avoid material knowledge "difficulties" : study of the effect on the BBBFJ appeared as an interesting compromise.

Literature about study of bolted elements representation [6], Bolted Flanged Joints (BFJ) [7], and even thermal shock effect on tightening in BFJ is quite plethoric [8], but they do not deal with the quantifying whether the dissymmetry of loadings induced by transient loading in a valve have any influence on the results or not.

## 1.3 The studied valve : Vanatome globe valve.

At the beginning of the NICODEME project, a scientific partnership was concluded with Vanatome S.A. who provided a globe valve and technical information. This is a globe valve of 100 millimeters nominal diameter. Figure 1 shows a partial vendor drawing of the globe valve with a photography of it mounted on EDF R&D facilities [1].

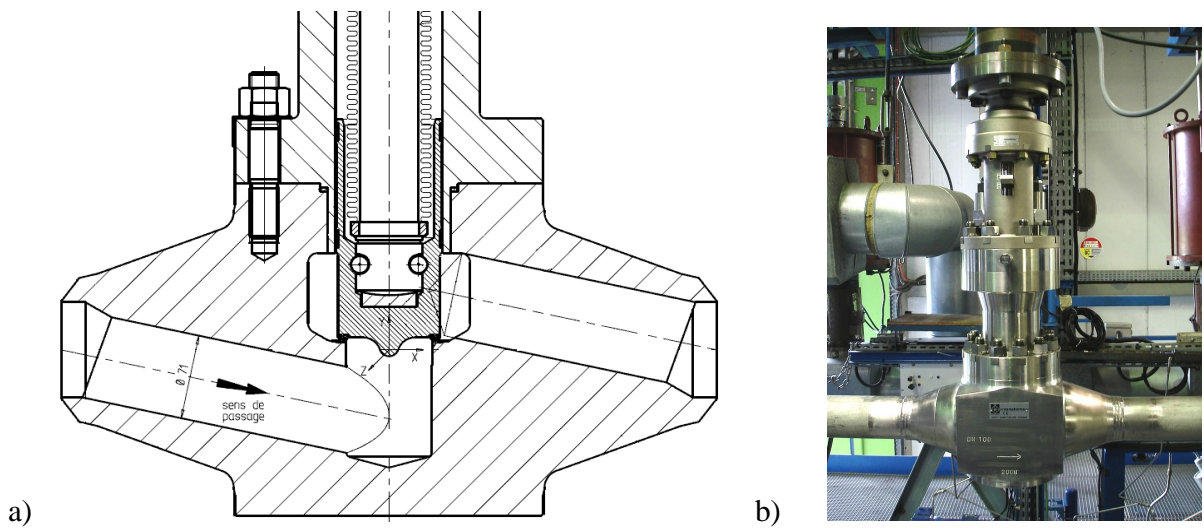

a) b)

**Fig. 1** : a) drawing of the body-bonnet flange joint (bolt is not in XY plane of symmetry, but is represented); b) picture of the valve at EDF R&D facilities.

# 2 Modeling Strategy

## 2.1 Mechanical representation of the valve

It was chosen to study the thermomechanical behavior of the complete valve, except for the actuator. This was considered as the easiest way to take into account the whole loading configuration of the valve. Previous work [9] showed difficulties that arise when one simplify the study by not considering the highest part of the valve : embedding Boundary Conditions (BCs) on the stem causes mechanical artifacts. One can find a complete description of the methodology used for these calculations in [10]. EDF own open-source Finite Element Method (FEM) code Code_Aster [11] was used for thermomechanical computations.

### 2.1.1 Meshing and representation considerations

Particular attention was taken to make a mesh that fulfills these conditions :

- Overall size small enough to run a one-step elastic calculation on a standard engineering workstation in about 1 hour.
- Adaptation of the mesh refinement to the supposed loadings.
- Creation of the topological entities adequate to the supposed BCs and of a clear list of them (e.g. creation of both surface used for contact, volumes to represent different parts…).
- Simplification of technological features considered as useless for envisaged computation analysis.

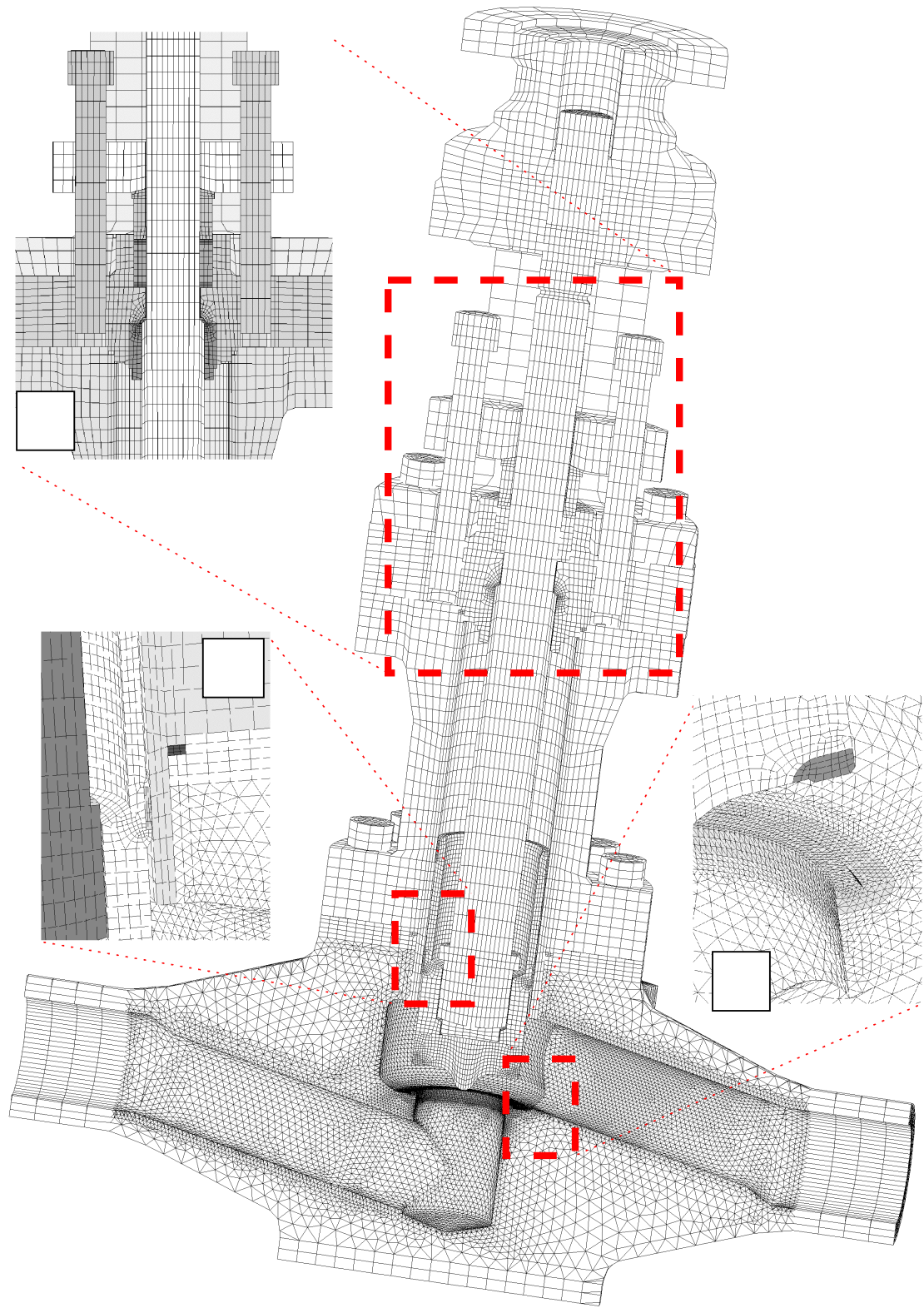

**Fig. 2** : Meshing of the globe valve. Some zone of interest are zoomed in : a) packing assembly, b) body-bonnet joint, and c) hard-facing on the sealing internal surface.

It is worth noting that this part of the modeling work was thought as a critical one : mechanical computation has to be already done (or at least thoroughly considered…) to avoid mistakes on meshing aspects. One can observe the global "mechanical" mesh geometry on figure 2. Its main characteristics are the following:

- Number of nodes : 100000, Number of 3D elements : 215000
- Number of groups of volumes: 37
- Number of groups of surface: 60 used in BCs.

### 2.1.2 Materials and thermomechanical behaviors

As a first step, all materials were considered as mechanically perfectly isotropic elastic. Young's modules, Poisson's ratios and thermal expansion coefficients were taken as temperature dependent whenever data were available. Steel grades characteristics were considered as in accordance with the rules used in the French nuclear industry (RCC-M French code [12]). Other materials characteristics (packing stuffing, hard-facing alloys…) were estimated from available literature. One can remark that pure elasticity is known to be a limitation for interpretation of results for some aspects [12,4].

### 2.1.3 Loadings, boundary conditions and temporal considerations

BCs are to be separated between "constant" ones (embedding to the reference, contacts between parts) and "temporal" ones (tightening of bolts, pressure appliance, thermal loadings).
Contacts were generally considered as friction-free, since no significant movement between parts was studieed (e.g. no stem operation). Two contact modelings were adopted :

- "simplified" contact modeling, not allowing surfaces normal relative displacement, which has the advantage of computational lightness.
- "full" unilateral contact modeling, using the active stresses method [16], and allowing a realistic contact modeling at higher computation cost.

Full contact modeling was attributed to surfaces of flange joints and surfaces subjected to potential interpenetration contradictory with design.

Nuts and threaded rods assembly were simplified to a "bolt-like" element (merging rod and nut in a single volume) which was considered as a good way to represent the technological function [17]. Threads are not represented, but tightening consists in relative displacement along threaded rod axis of 2 linear groups of nodes belonging respectively to the threaded piece and to the rod, and representing the first tight threads. This is schematized in figure 3. Tightening of rods (i.e. displacement $s$) was adjusted within one iteration to fulfill valve specifications.

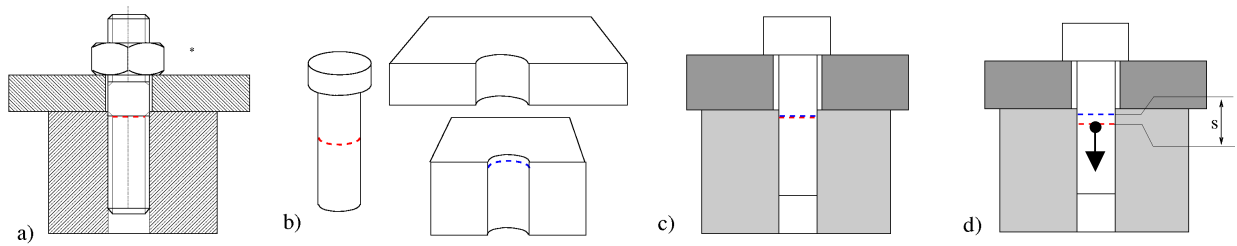

**Fig. 3** : Representation of the tightening by threaded rods and nuts in this study. a) drawing ; b) groups of volumes ; c) initial positioning ; d) tightening of the rod by relative movements of nodes

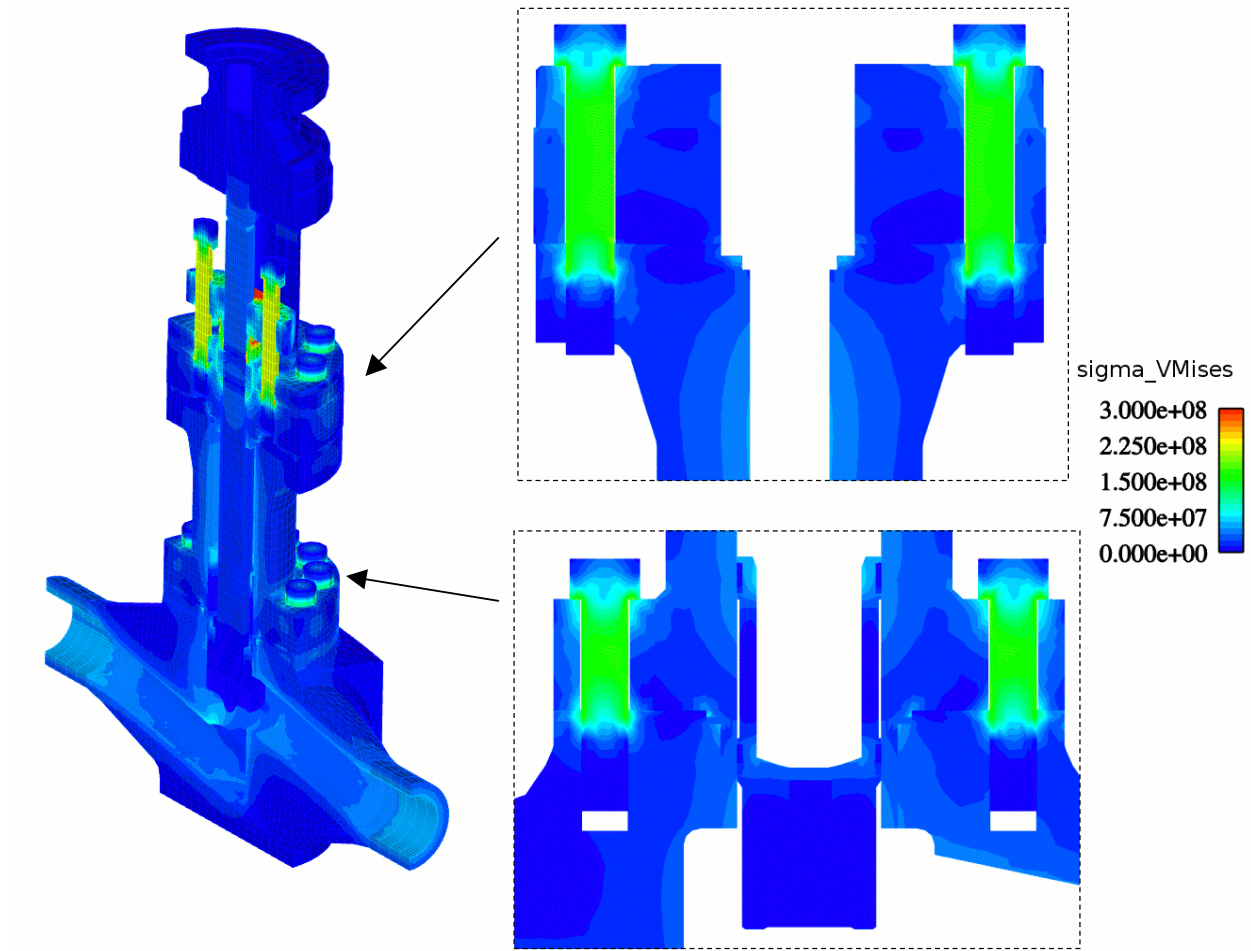

**Fig. 4** : Stress repartition inside the valve at its initial state after the assembly sequence.

Study of the loading was cut in sequences that allowed separating and analyzing each aspects of the valve service:

- mechanical assembly (mainly tightening of threaded elements),
- mounting on the test loop (stabilized temperature equilibrium, internal pressure, and end effect application)
- transient loading (thermal or pressure shock)

These sequences can be separately computed, so that assembly and mounting sequences can be used for pressure or thermal shock computation. One can observe the Von Misses stresses inside the loaded parts of the valve after the assembly sequence on figure 4.

## 2.2 Thermal shock modeling

### 2.2.1 Modeling principles

The thermal shock to represent was chosen to be representative of experimental conditions applied to the globe valve during functional tests. This consists of applying a thermal shock between 60°C and 285°C, valve wide open, with an internal fluid pressure of 185 bars. Thermal insulation will cover lower part of the valve from the body to half of the bonnet (i.e. all the volume represented in figure 1.a will be insulated).

As for mechanical representation, it was chosen to make the best representation possible of the thermal gradient during the shock. This is expected to give a precise description of thermomechanical state inside the valve, and will possibly allow an *a posteriori* simplification of thermal BC applied in future modeling, with ability to evaluate whether simplification influences results or not.

Two additional meshes were done using the same geometry as for the mechanical mesh. The first one represents the inner fluid volume to be used for Computational Fluid Dynamics (CFD) calculations, and the second one is a simplified representation of the valve volumes, used for thermal calculation. A particular attention was paid to the concomitant interfacial surface, which needs to be geometrically the same in both meshes.

This surface will be the interface for coupling CFD and thermal calculation through calculation of the thermal exchange between fluid and solid at each time step. Computation of each thermal shock (i.e. cold shock when fluid temperature is changed from 285°C to 60°C and hot shock in reverse case) were made on a cluster of 64 Dual-core Itanium2™ for 2 weeks. It is worth noting that only the CFD code (EDF's Code_Saturne [13]) supports parallelism, so that the thermal code (EDF's Syrthes [14]), which does not support it, used only one core. More precisions about fluid-thermal computation methods used for this simulation can be found in [18].

### 2.2.2 Thermal shock modeling results

Figure 5 presents the CFD results obtained at the beginning of a cold shock: the thermal field inside the fluid resulting of the shock and thermal exchanges at the interfacial surface are presented. This last field will be used as a thermal BC for thermal simulation of the transient effect. Figure 6 presents thermal field inside the valve volume during both thermal shocks. One should remark that a second time scale ($t2$) has been introduced. It will be used to compare hot and cold shock on the same time basis. Specific instants of interest are also introduced. H0 and C0 will stand respectively for the hot shock beginning state (cold equilibrium) and cold shock beginning state (hot equilibrium).

H2 and C2 stand for times of higher "transient" mechanical effect for both hot and cold shock (they are not known *a priori*, but are results of the mechanical study part). One can notice by comparing times for both $t$ and $t2$ scales that cold shock is slightly quicker than the hot one. This is to be explained by the effect of the higher part of the valve that is considered in air at room temperature, and act as a heat bridge. This phenomenon supports cooling and on the other hand opposes heating.

## 2.3 Thermo-mechanical computations

### 2.3.1 Thermal shock modeling projection

Results of the thermal computation were imported inside Code_Aster FEM Solver, and temperature fields projected on the mesh designed for thermomechanical study. Computation of one time step took about 1 hour on a single workstation. Parallelism was not used since it won't be useful in a problem involving unilateral contact.

As the problem has a symmetry on the XY plane, only half of the valve will be modeled (the one with Z<0). Consequently, only 6 (among 12) threaded rods will be included in the modeling of the BBBFJ.

### 2.3.2 Transient description issues

Each thermal shock took about 30 time steps to be mechanically "correctly" described. It is worth noting that "correct description" of mechanical fields during such thermal transients consists of choosing the time steps adequately so that they allow representing local maximums of the value of interest on a temporal point of view. This aspect is negligible when dealing with "reasonable" quantities of data, but becomes critical when computational needs reaches hardware limitations. This implies that the process is iterative, since you don't know *a priori* the temporal evolution of the result, and you can't afford to multiply time steps.

The same issue arises when trying to determine the transient duration. This was treated by calculating the thermal field at equilibrium separately, and stopping computation when thermal transient approaches the equilibrium state.

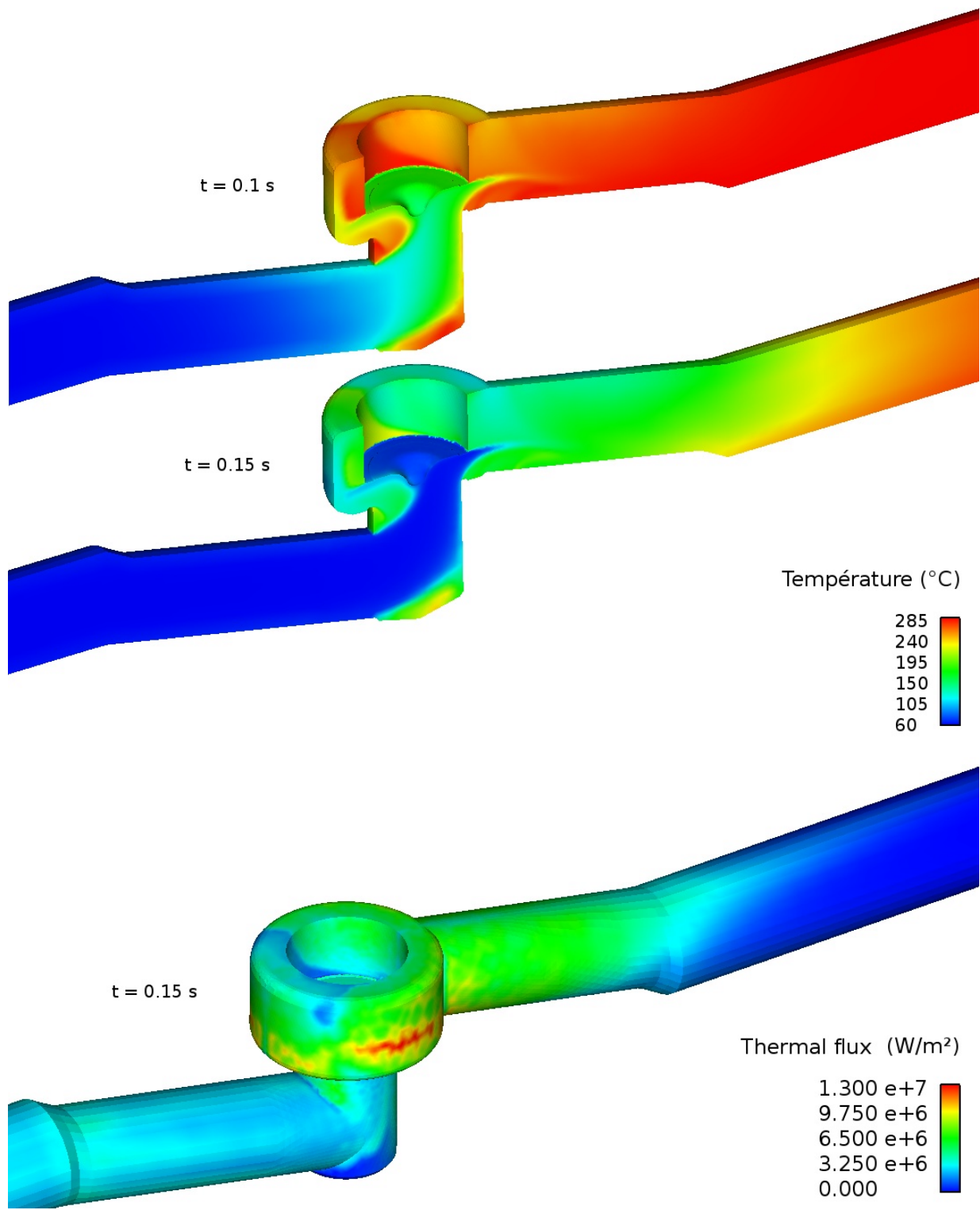

**Fig. 5** : CFD results. Temperature inside fluid at the beginning of a cold shock and resulting thermal flux (solid to fluid).

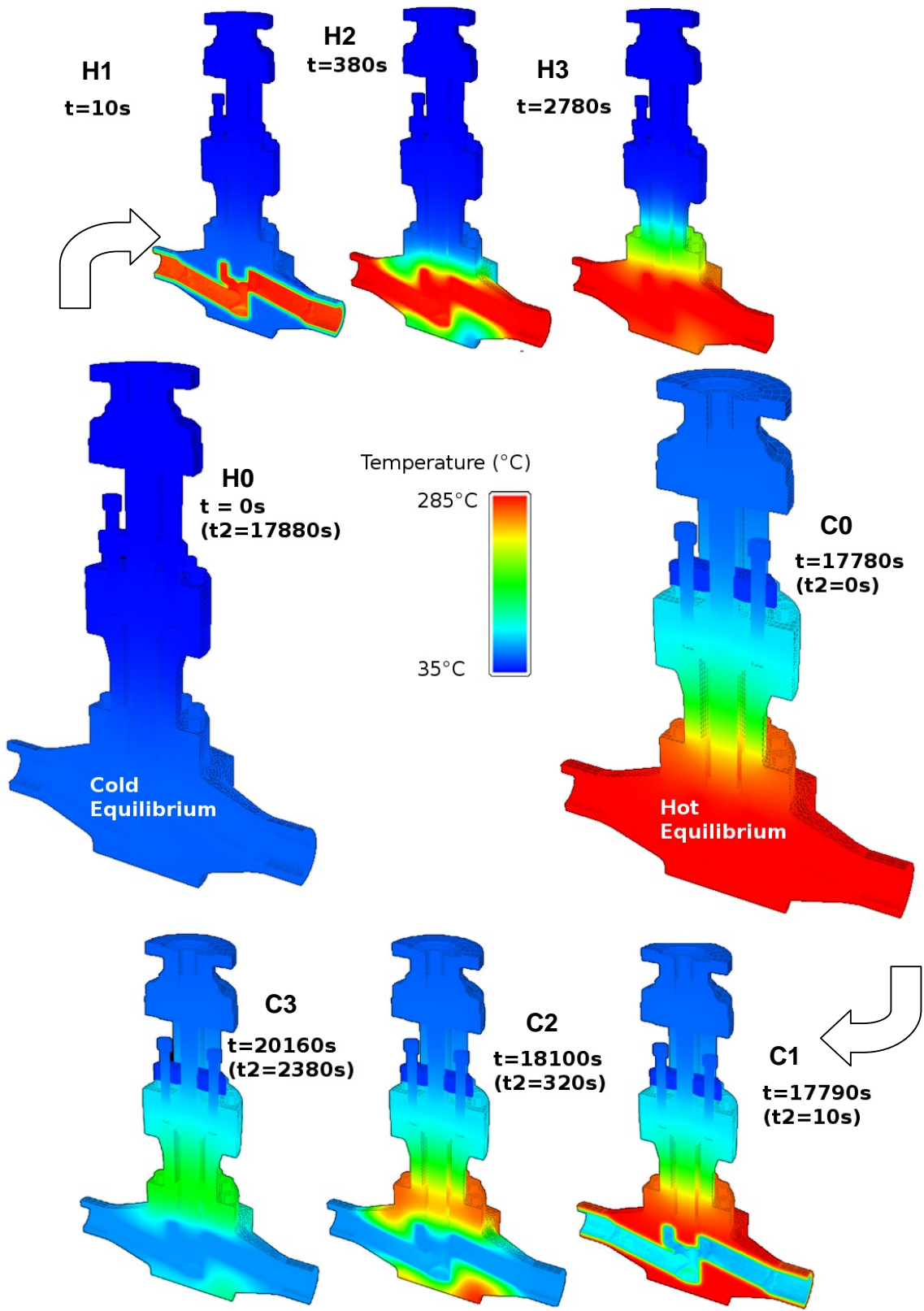

**Fig. 6** : Thermal computation results. Temperatures maps during a full alternate shock

# 3 Thermal shock effect on the body-bonnet bolted flange joint

## 3.1 Mechanical response of the threaded rods.

Threaded rods where chosen as mechanical sensors during the thermal shock since they are believed to fulfill two important features :

- They are believed to be sensitive to the thermal transient effects
- They are (relatively) easy to fit with instrumentation

This part will discuss the mechanical response of the BBBFJ with special attention on the tightening of the rods. Other aspects (e.g. gasket tightening aspects) won't be discussed in this work, but they will be once experimental validation of rods efforts described in paragraph 4 will be achieved.

Thermomechanical behavior of BFJ are usually dimensioned and designed as standards piping bolted flange junctions. It is worth noting that "standard" BFJ will now be used to denominate an axial flanged tube connection with metallic gasket and clearance between the flanges. A complete study of these junctions where made within [20]. Main conclusion of this work were the following :

- During hot shock, tightening elements (threaded rods, screws) are submitted to a transient over-tightening.
- During cold shock, they are submitted to a transient under-tightening.

These variations are generally considered to be almost symmetric between the two shocks : absolute values of the over and under-tightening being almost the same. These transient variations are also generally considered as homogeneous on every tightening element of the junction (not considering primary tightening dispersion due to the assembly process, which is not discussed hereby).

Figure 7 presents the transient tightening results obtained during an alternate thermal shock. Resultant efforts inside bolts are presented for the whole transient, as well as details of the beginning of both hot and cold shocks. Instants of interest introduced in figure 6, and the initial tightening efforts (obtained after assembly sequence at 20°C) are also represented.

**Fig. 7** : a) Tensile force inside rods during alternate shocks. b) Beginning of hot transient and, c) beginning of cold one. Instants of interest are also noted on b) and c).

## 3.2 Results comments

### 3.2.1 Un-tightening at Thermal equilibriums

The first results concerns the general under-tightening that is observed when the valve reaches both equilibrium (instants H0 and C0). Explanation of these under-tightening can be found in the fact that the assembly sequence is made at 20°C, and that cold and hot equilibrium stands for fluid at respectively 60°C and 285°C. Thermal coefficients between rods and bonnet are about the same for these considered temperature so that such an under-tightening is not due to the use of different materials.

Figure 8 shows the limit of the insulation of the lower part of the valve, but also thermal fields of the bonnet and two rods (1 & 6) for both equilibriums with appropriate scales (upper part of the bonnet colors are thus not significant).

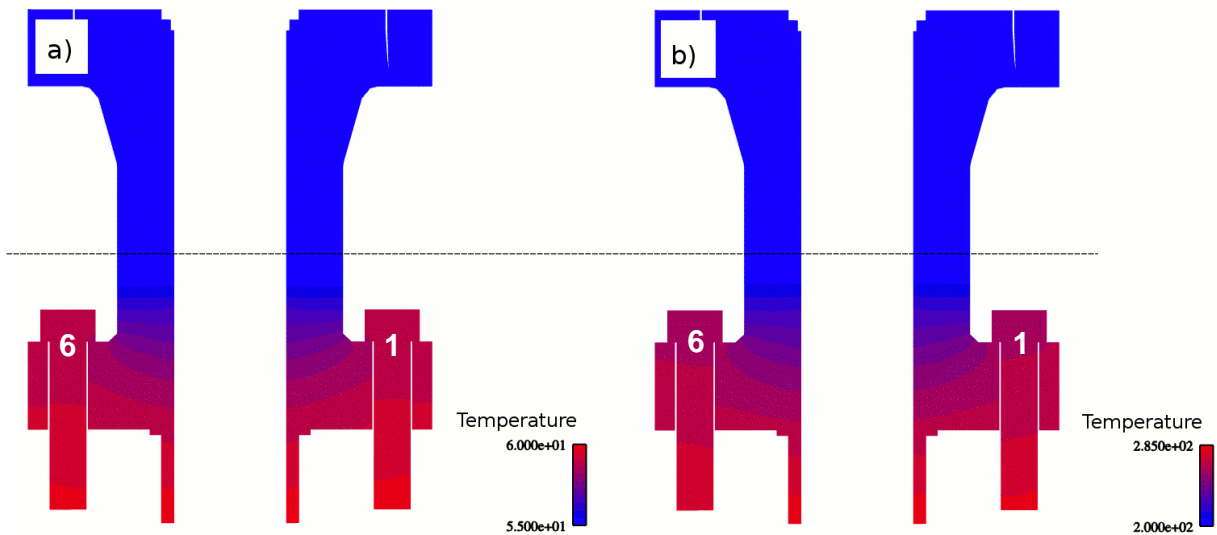

**Fig. 8** : Temperature (°C) of bonnet and bolts (1, 6) : a) cold equilibrium, b) hot equilibrium. Dashed line represents the limit of thermal insulation of the valve (temperature above his line are lower than the represented minimum).

One can easily see that there is a slight average difference in temperature between the tightened volume (bonnet lower flange) and the tightening element (rods), so that the rod is a little bit hotter. This explains the slight under tightening during both equilibriums, and its intensity difference can be observed by looking at the temperature scales : for hot equilibrium, difference in temperature is more important.

It is worth noting that even for the lowest tightening effort during the hot equilibrium and transient, efforts developed by threaded rods (~792 KN) are sufficient to fulfill the minimum effort to ensure leaktightness accordingly to French standards (691KN, [19]). No relevant differences between rods tightening forces can be observed during these equilibriums. The other point that deserves attention concerns the thermal insulation.

One can easily envisage that a different insulation configuration could lead to different results.

Although thermal field heterogeneities intensity explains under-tightening in stabilized thermal states, it does not allow understanding thermal transient load variations.

### 3.2.2 Thermal transients global effects : over-tightening during both shocks

Figure 9 and 10 show the evolution of behavior at the interface between the two tightened elements, with particular attention to the contact surface clearance. The presented field corresponds to the von Mises equivalent stress, signed by the hydrostatic pressure :

$$\sigma_{VMIS\_SG} = \frac{tr(\underline{\sigma})}{|tr(\underline{\sigma})|} \cdot \sqrt{\frac{1}{2}((\sigma_{xx} - \sigma_{yy})^2 + (\sigma_{yy} - \sigma_{zz})^2 + (\sigma_{zz} - \sigma_{xx})^2}$$

This represents the stress state (positive for tensile stress state, negative for compressive one), but also its intensity.

During the thermal transients, flanges will undergo non-uniform thermal field that will result in a certain amount of bowing of strait profiles. Particularly, the contact surfaces of the flange will become deformed, which can be observed by observing the clearance profiles during the shocks. These deformation have the same tendencies that the one that were previously described by Scliffet [20] on standard BFJ. Global deformation leads to different states :

- For hot shock, transient thermal deformations conduct to a bowing geometry so that flanges contact mainly takes place in the internal part of the flange.
- For cold shock, transient thermal deformations conduct to a bowing geometry so that flanges contact mainly takes place in the external part of the flange.

BBBFJ bolts mean tightening behavior during shocks obviously differs from the one that would have been expected when considering a "standard" BFJ. The design of the BBBFJ (which features a "full-face" contact of flanges) is somehow different from a standard BFJ. Figure 11 schematically illustrates the fact that design of the BBBFJ has no initial clearance between flange surfaces. This difference explains the over-tightening that takes place in both hot and cold transient.

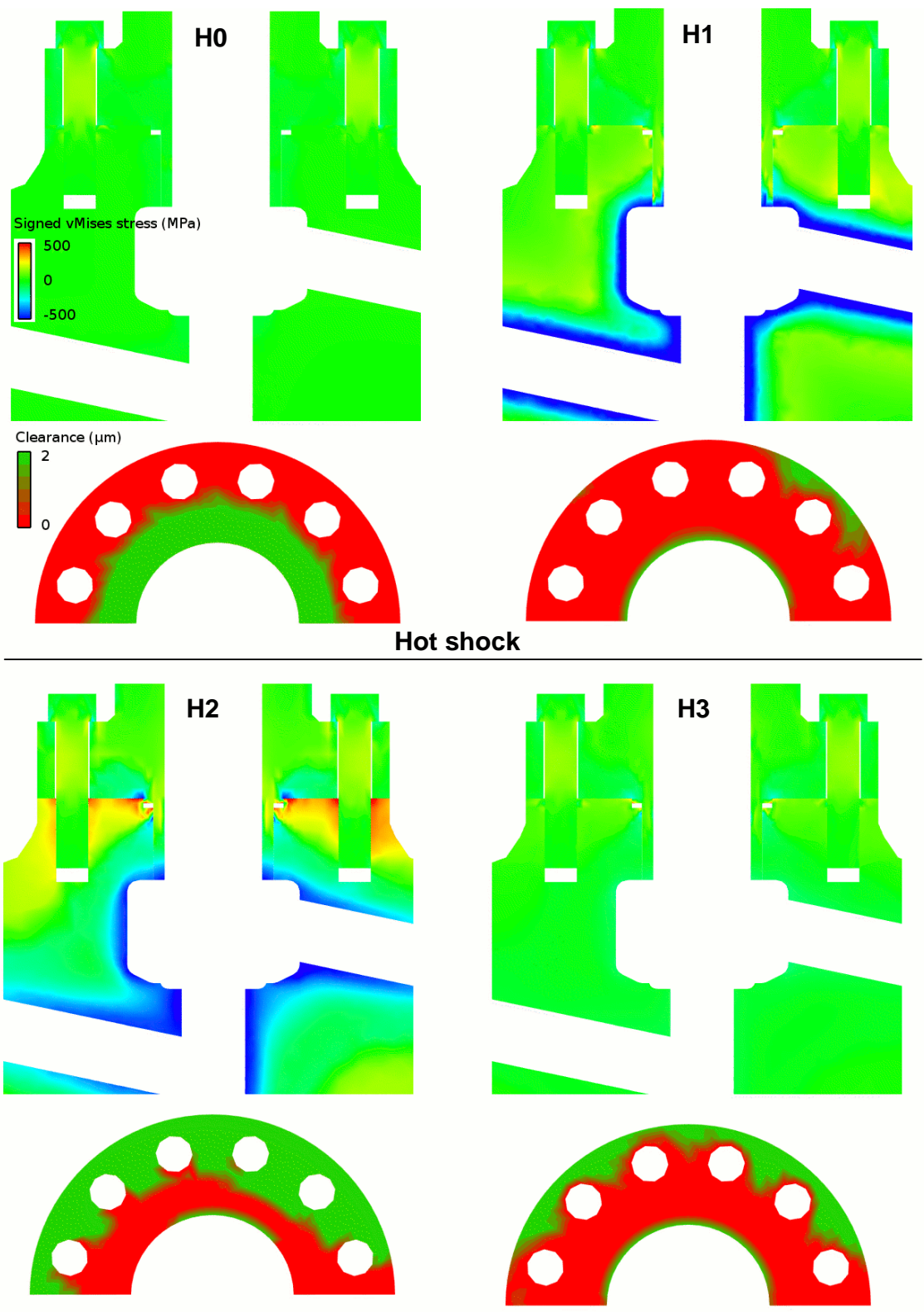

**Fig. 9** : Evolution of signed von Mises stress inside the BBBFG and of clearance between flange surfaces during hot shock.

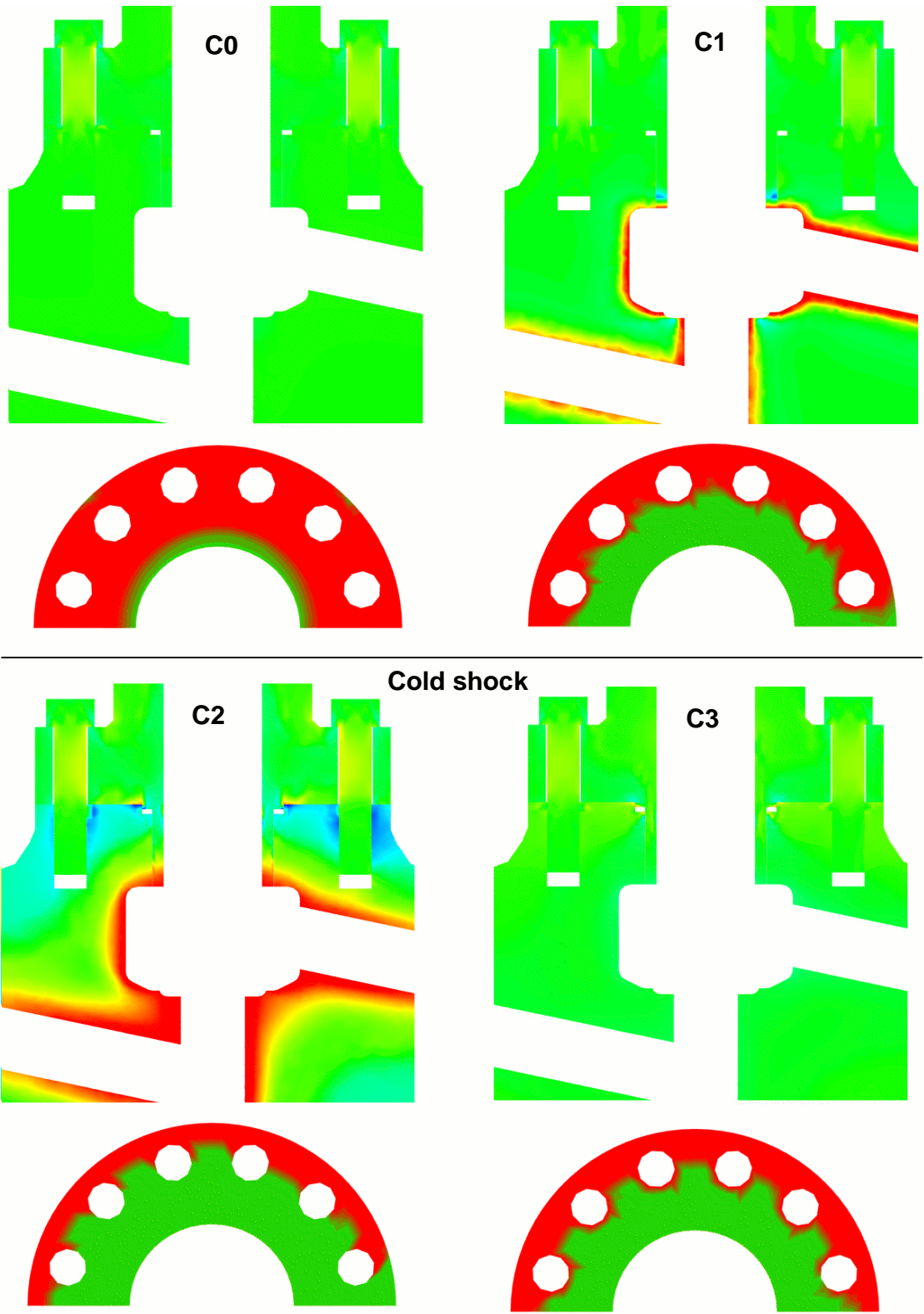

**Fig. 10** : Evolution of signed von Mises stress inside the BBBFG and of clearance between flange surfaces during cold shock. Scales are indicated of Figure 9.

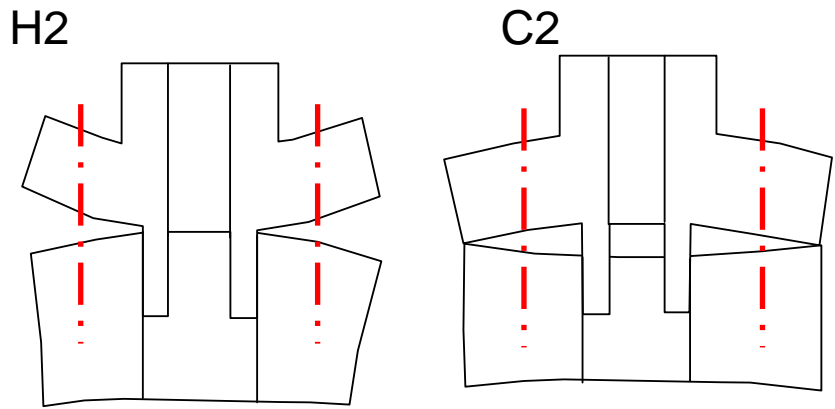

**Fig. 11** : Schematic representation of deformed shapes of volumes of BBBFJ for moments of interests related to highest over-tightening.

On a standard BFJ, a clearance is set so that only the gasket surface will be used as an interface. This is mainly due to the fact that bolts tightening torque control tightening. This implies that a clearance already exists that is too large to be compensated by flanges global deformation. This BBBFJ has another design : the initial dimension of the seal housing mainly controls gasket tightening, and rods tightening are dimensioned to ensure tightness.

The hereby "mechanical" analysis does not take into account another quite important difference between this BBBFJ and standard BFJ, which is represented in this work : on the thermal point of view, heat exchanges are quite different between the two configurations. When a standard BFJ usually undergoes "radial" thermal exchanges (relatively to flange axis), heat transfer mainly happens along the flange axis direction in a valve. One should also remark that unilateral contacts used as BCs for flanges surfaces contact is quite important, since simplified contact would have led to bad interpretation of tightenings evolution, by hiding the bowing effects.

### 3.2.3 Thermal transients local effects: non-intuitive variations among threaded rods

Another difference can be found between the behavior of BBBFJ : thermal field can't be considered as axi-symmetric relatively to the flange axis. This creates differences in behavior between the rods during shocks. Tightening relative difference between rods can reach 34% of the recommended initial tightening force (25 KN compared to 72 KN between rods 1 and 2 at the instant H2, fig 7.b). These differences in behavior are also supposed to be dependent of the thermal strain repartition.

One should remark that experiments are thus required to check whether modeling thermal hypothesis are relevant or not. E.g., thermal conduction between flanges surfaces is supposed to be perfect, not taking into account either the thermal resistance of the interface nor the fact that the clearance gap that appears can disturb thermal exchanges.

# 4   Discussion

## 4.1 Modeling results

Thermomechanical unusual effects on the BBBFJ are predicted by the global modeling in this work when comparing results with previous work made on standard BFJ. These effects are the following:

- Hot thermal equilibriums (with fluid at an higher temperature than ambient temperature used for assembly) lead to global under-tightening of threaded rods. No relevant difference between rods being found.
- Contrary to usual BFJ, both hot and cold transient leads to over-tightening of the threaded rods. Hot transient leading to the higher over-tightening. This difference is attributed to the "full-face" contact design of the BBBFJ. During shocks, over-tightening is quite heterogeneous between rods.

## 4.2 Forthcoming experimental validation

These results could be a basis for a global sealing performance study of the BBBFJ. But they have to be validated before to be used in comprehensive analysis. The valve used in this study will be instrumented in order to check the thermal part of the modeling but also to dynamically follow the threaded rods tightening. Figure 12 shows the experimental set-up that is currently mounted on the rods and on the body of the studied valve.

Thermocouples will be used to check that the thermal modeling is correct and strain gages will allow evaluating strain (i.e. tightening efforts variations) inside the whole threaded rods ring of the BBBFJ. Initial assembly will be done with measurements so that the whole rods tightening history can be taken into account. Several hypothesis that were made in this work will thus be confirmed or contradicted, allowing to improve iteratively the modeling, and furnishing a better comprehension of the relevant conditions to fulfill when modeling thermal shocks on flanged connections.

## 4.3 Further work

Once validation of the thermomechanical behavior will be achieved and experimentally validated, a lot of work in various aspects remains to allow a numerical validation of the BBBFJ sealing during thermal shocks. The relation between leakage, pressure and local internal constraint are quite empirical. Giving a better description of gasket materials behaviors, and taking into account a possible relaxation of tightening due to plasticity or creep that occurs at threaded roots could allow a better diagnostic of in-service leakage, which also requires progress in the sealing performance of surfaces evaluation field [21].

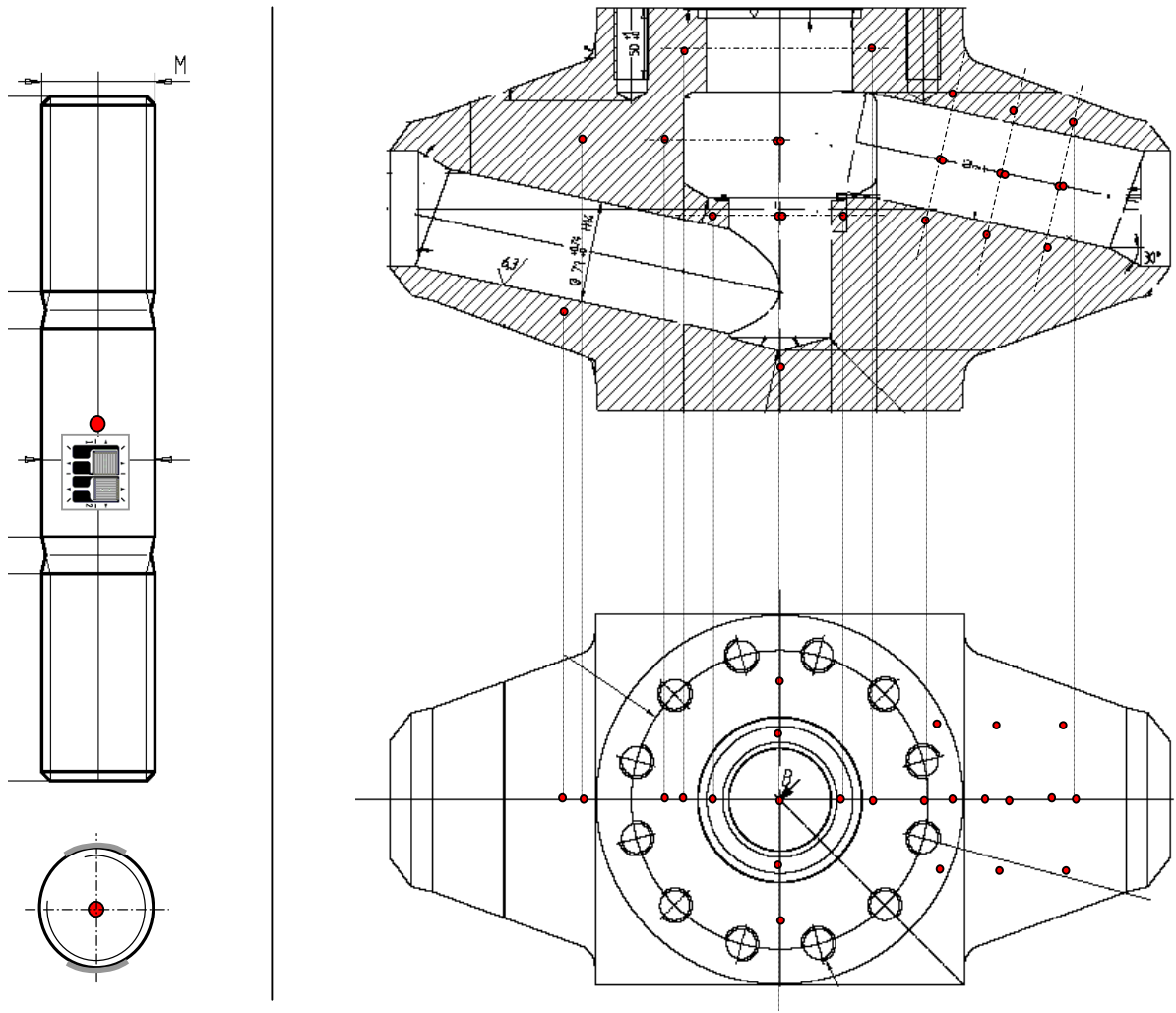

**Fig. 12** : Positioning of the thermocouples (red dots) and of strain gages (Vishay™ WK gages) used for experimental validation for the threaded rods and the body. Thermocouples will also be implemented in other valve parts

# 5 Conclusion

Thermal shock effects on the BBBFJ behavior of a valve used in nuclear industry were evaluated using numerical simulation. Choices were made about modeling that lead to the biggest "reasonable" calculation available in a standard engineering office, once CFD results are available. Modeling reveals several noticeable differences between thermomechanical behavior of the BBBFJ compared to a standard BFJ, attributed to design features and differences between thermal loadings.

All these reported differences will be validated or contradicted in further experimental work, which will allow iterative progress in the modeling part, and objective criticism of the modeling choices.

Would the modeling happen to be good "on first shot", much work remains to do to numerically assess complex technological functions such as sealing of BFJ during transient loadings. Future work on this valve will focus on the evaluation of modeling results by comparison with experimental ones, and its possible use for other applications, like hard-facing thermal shock strength evaluation.

# 6  Acknowledgements